\documentclass[aps,prb,twocolumn,groupedaddress,showpacs]{revtex4}

\usepackage{graphicx}
\usepackage{color}
\usepackage{epstopdf}
\usepackage{epsfig}
\begin{document}

\title{Circuit approach to photonic heat transport}

\author{L. Pascal$^{1}$, H. Courtois$^{1}$ and F. W. J. Hekking$^{2}$}
\affiliation{$^{1}$Institut N\' eel, CNRS and Universit\'e Joseph Fourier, 25 Avenue des Martyrs, BP 166, 38042 Grenoble, France.\\
$^{2}$LPMMC, Universit\'e Joseph Fourier and CNRS, 25 Avenue des Martyrs, BP 166, 38042 Grenoble, France.}

\date{\today}

\begin{abstract}
We discuss the heat transfer by photons between two metals coupled by a circuit containing linear reactive impedances. Using a simple circuit approach, we calculate the spectral power transmitted from one metal to the other and find that it is determined by a photon transmission coefficient, which depends on the impedances of the metals and of the coupling circuit. We study the total photonic power flow for different coupling impedances, both in the linear regime, where the temperature difference between the metals is small, and in the non-linear regime of large temperature differences.
\end{abstract}

\maketitle

\section{Introduction}
Electron thermodynamics at the nano-scale is a fast developing topic,\cite{RMP-Giazotto} in particular in superconductor-based hybrid devices. For instance, Superconductor-Insulator-Normal (S-I-N) metal junctions biased just below the superconducting energy gap display electronic cooling.\cite{RMP-Giazotto,PRL-Rajauria} The thermal properties of S-N and S-N-S hybrid devices also show signatures of quantum phase coherence.\cite{SST-Chandrasekhar}

In metallic systems, heat conduction can be achieved by electrons, phonons and also photons.\cite{Johnson-Nyquist,PRL-Schmidt} The photonic channel was recently revealed experimentally \cite{Nature-Meschke,PRL-Timofeev-1} at very low temperature in devices including superconducting transmission lines. In a superconductor, electrons are paired into Cooper pairs so that the electron-phonon coupling is vanishing,\cite{PRL-Timofeev-2} just as the electronic heat conductance. Therefore, only photons can contribute to the heat transfer at very low temperature. With a good matching between the source and the drain, the conductance of a superconducting transmission line is equal to the thermal conductance quantum:\cite{Pendry_83} $K_Q =  k_{B}^{2}T\pi / 6 \hbar$.

The photonic channel for heat transfer can in principle couple metallic systems that are galvanically isolated, e.g. through a capacitor. This effect can be beneficial in some cases, but also detrimental when one wants to maintain two electronic populations at different quasi-equilibrium temperatures. An inductance can also be present in some realistic configurations, due to the wiring geometry. In this paper, we investigate the photonic heat transfer through a general reactive impedance, i.e. a linear coupling circuit that contains a capacitor, an inductance, a resonant circuit or a transmission line. We follow a simple circuit approach, valid at low temperatures when the relevant photons have wavelengths larger than the size of the typical circuit element. The metallic parts can then be treated as lumped elements characterized by an electrical impedance. We present a quantitative analysis, enabling us to establish design rules for useful devices, including phonon thermometers or electron coolers.

\section{Circuit approach}

\begin{figure}
\centering
\includegraphics [width=8.5 cm]{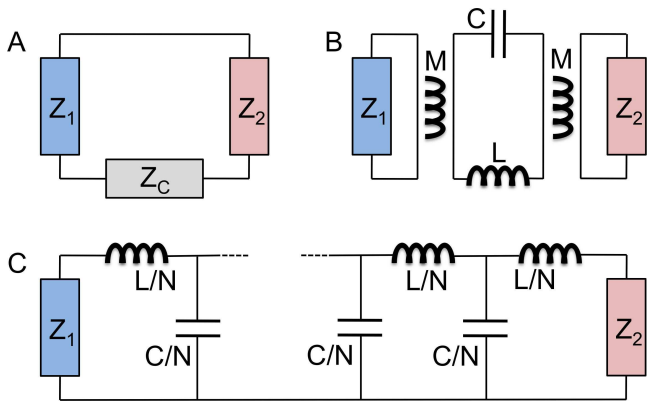}
\caption{Top left panel, circuit A: two impedances $Z_1$ and $Z_2$ are coupled by a purely reactive coupling impedance $Z_c$. Top right panel, circuit B: two impedances $Z_1$ and $Z_2$ are coupled via mutual inductances to a purely reactive central circuit of tota impedance $Z_c$ (here an LC resonator). Bottom panel, circuit C: two impedances $Z_1$ and $Z_2$ are coupled via a transmission line.}
\label{Circuits}
\end{figure}

We consider the circuits A, B and C shown in Fig.~\ref{Circuits}. These configurations contain two impedances $Z_i(\omega)$ ($i=1,2$), kept at different temperatures $T_i$ such that $T_2 >T_1$. We wish to analyze heat flow between these elements driven by the temperature difference only, i.e. in the absence of any voltage or current source. In configurations A and C, the two impedances belong to {\em the same} circuit and are coupled through a purely reactive coupling element with impedance $Z_c(\omega)$ or a transmission line. Configuration A is a generalization of the one considered in Ref.~\onlinecite{PRL-Schmidt}. In configuration B, the two impedances belong to {\em two different} circuits, both coupled via a mutual inductance $M$ to a third linear circuit that mediates the heat transfer. Again we assume this coupling circuit to contain reactive elements only with a total impedance $Z_c(\omega)$. Configuration B was analyzed in Ref.~\onlinecite{PRL-Jauho} with non-equilibrium Green function techniques; here we will show that our circuit approach yields the same results. To the best of our knowledge, configuration C has not been analyzed until now.

In the absence of voltage or current sources, electromagnetic fluctuations are responsible for the heat flow between impedances. We therefore start our analysis by analyzing the current and voltage fluctuations induced by the various circuit elements. Following Ref.~\onlinecite{Beenakker_03}, we decompose the fluctuating current $\Delta I_i$ through the $i$th element into two parts:
\begin{equation}
\Delta I_i=\delta I_i+(1/Z_i) \Delta V_i.
\label{defin}
\end{equation}
The quantity $\Delta V_i$ is the voltage fluctuation across the element $i$. The fluctuation $\delta I_i$ is the intrinsic fluctuation produced by the element due to Johnson-Nyquist noise with spectral function
\begin{equation}
\langle \delta I_{i}(\omega)\delta I_{i}(\omega') \rangle = 2\pi\delta(\omega+\omega') C_{i}^{(2)}
\label{didi}
\end{equation}
where $\langle ... \rangle$ denotes a thermal average and
\begin{equation}
C_{i}^{(2)} =  \hbar \omega Re[1/Z_i(\omega)] \coth(\frac{\beta_i \hbar \omega}{2}),
\label{JN}
\end{equation}
with $\beta_i = 1/k_B T_i$ being the inverse temperature of element $i$. The $\delta$-function in Eq.~(\ref{didi}) reflects the fact that we consider noise in the stationary limit; Eq.~(\ref{JN}) is the fluctuation-dissipation theorem,\cite{Landau_Lifschitz} written in a form appropriate for Johnson-Nyquist noise generated by an impedance $Z_i$, kept at an inverse temperature $\beta_i$.

Certain constraints apply to the fluctuations $\Delta I_i$ and $\Delta V_i$. For instance in circuit A, current conservation implies $\Delta I_i = \Delta I_j \equiv \Delta I$ for any element $i,j = 1,2,c$; similarly, $\sum _i \Delta V_i =0$. We assume these constraints to be satisfied simultaneously for each element; this implies that we consider fluctuations at low frequencies $\omega$ such that the wavelengths $\lambda_\omega \propto 1/\omega$ of the relevant electromagnetic waves are larger than the typical size of the circuit. At 1 Kelvin, this corresponds to a maximum circuit size of about 1 cm; this limit scales as $1/T$.

We will be interested in the net power $P_{i}(t)$ absorbed by the element $i=1,2$ as a function of the temperature difference $\Delta T = T_2-T_1$ between them. This quantity is given by $P_{i}(t) = \langle \Delta I_{i}(t) \Delta V_{i}(t) \rangle$. It can be expressed in terms of the frequency-dependent correlation function $\langle \Delta I_{i}(\omega)\Delta V_{i}(\omega') \rangle$:
\begin{equation}
P_{i}(t) =\int\frac{\emph{d}\omega}{2\pi} \int\frac{\emph{d}\omega'}{2\pi} e^{-i\omega t}e^{-i\omega' t} \langle \Delta I_{i}(\omega)\Delta V_{i}(\omega') \rangle.
\label{Pi}
\end{equation}
We will calculate and analyze this quantity in detail below for circuits A, B and C.

\section{Calculation of the heat exchange}

\subsection{Heat exchange with direct coupling}

Let us first consider circuit A in Fig.~\ref{Circuits}, for the simple case where the reactive coupling element is absent, $Z_c(\omega) = 0$. Imposing the constraints on $\Delta I_i$ and $\Delta V_i$, we obtain $\Delta V_1(\omega) = -\Delta V_2(\omega) = Z_1(\omega) Z_2(\omega)[\delta I_2(\omega) - \delta I_1(\omega)]/[Z_1(\omega) + Z_2(\omega)]$. Using Eq.~(\ref{didi}) and the fact that the current noise in different elements is uncorrelated, $\langle \delta I_1 \delta I_2\rangle = 0$, we find after some elementary algebra,
\begin{eqnarray}
\langle \Delta I_{1}(\omega)\Delta V_{1}(\omega') \rangle = -\langle \Delta I_{2}(\omega)\Delta V_{2}(\omega') \rangle = \nonumber \\
2 \pi \delta(\omega + \omega')
\frac{Z_1(\omega') Z_2(\omega')}{Z_t(\omega) Z_t(\omega')}  \nonumber \\
\times [Z_2(\omega) C_2^{(2)}(\omega) - Z_1(\omega) C_1^{(2)}(\omega)].
\label{corr}
\end{eqnarray}
Here $Z_t(\omega) = Z_1(\omega) + Z_2(\omega)$ is the total series impedance of the two elements. Substituting Eq.~(\ref{corr}) into Eq.~(\ref{Pi}), using Eq.~(\ref{JN}) and integrating over $\omega'$, we find
\begin{eqnarray}
P_{1}(t) = -P_2(t) = \nonumber \\ \int_{-\infty}^{+\infty}\frac{\emph{d}\omega}{2\pi} \frac{2 \hbar \omega Re[Z_{1}(\omega)] Re[Z_{2}(\omega)]}{\mid Z_{t}(\omega)\mid ^{2}}[n_{2}(\omega) - n_{1}(\omega)],
\label{centrres}
\end{eqnarray}
where we made use of the relation $Z(-\omega) = Z^*(\omega)$ as well as of the identity $\coth x = 1+ 2 n(x)$, with $n (x) = [e^x-1]^{-1}$. The result reflects energy conservation: the power emitted by one of the elements is absorbed by the other.

If we define an effective photon transmission coefficient
\begin{equation}
{\cal T}(\omega) = \frac{4 \Re\mbox{e}[Z_{1}(\omega)] \Re\mbox{e}[Z_{2}(\omega)]}{\mid Z_{t}(\omega)\mid ^{2}},
\label{centrrescoef}
\end{equation}
and use the fact that the integrand is an odd function of $\omega$ to restrict the integration to positive frequencies only, we can write:
\begin{equation}
P_{1}(t) = -P_2(t) =  \int_{0}^{\infty}\frac{\emph{d}\omega}{2\pi} \hbar \omega {\cal T}(\omega)[n_{2}(\omega) - n_{1}(\omega)].
\label{centrressimpl}
\end{equation}
Note that this expression is similar to the one used to discuss heat transport in electron~\cite{sivan_86} and phonon systems~\cite{angelescu_98} within the scattering approach. The net heat current is obtained as the difference of the two heat-currents emanating from impedances 1 and 2, that serve as photonic reservoirs each characterized by an equilibrium Bose-Einstein distribution function at temperature $T_{1,2}$. Part of the total heat current injected by the photonic reservoirs can be reflected back, due to a mismatch between impedances $Z_1$ and $Z_2$. The amount of back-reflection can be frequency dependent; it is described by the frequency-dependent transmission coefficient ${\cal T}(\omega) \leq 1$.

Interestingly, as we will show below, Eqs.~(\ref{centrrescoef}) and (\ref{centrressimpl}) describe the heat transfer between impedances 1 and 2 {\em for arbitrary configurations A, B and C, as long as the coupling elements are purely reactive}. The corresponding total impedance $Z_t$ will be a configuration-dependent function of $Z_1$, $Z_2$ and the impedance of the coupling element.

We will be interested in the heat transfer between elements 1 and 2 as a function of the temperature difference $\Delta T = T_2-T_1$. For later use we define $\Delta \beta = \beta_1 - \beta_2$ and $\beta = (\beta_1 +\beta_2)/2$; then Eq.~(\ref{centrressimpl}) can be rewritten as
\begin{eqnarray}
P_{1}(t) = -P_2(t) =  \nonumber \\
\int_{0}^{\infty}\frac{\emph{d}\omega}{2\pi} \hbar \omega {\cal T}(\omega)\frac{\sinh \Delta \beta \hbar \omega/2}{\cosh \beta \hbar \omega - \cosh\Delta \beta \hbar \omega/2}.
\label{centrressimplbeta}
\end{eqnarray}

If the temperature difference $\Delta T = T_2-T_1$ is small compared to $T_1$ or $T_2$, we can expand the integrand to first order in $\Delta \beta$. The expression (\ref{centrressimplbeta}) then reads
\begin{equation}
P_{1,2}(t) = \pm \int_{0}^{\infty}\frac{\emph{d}\omega}{2\pi} \hbar \omega {\cal T}(\omega) \frac{\Delta \beta \hbar \omega/2}{\cosh \beta \hbar \omega - 1},
\end{equation}
and can be further simplified to:
\begin{equation}
P_{1,2}(t) = K \Delta T = \frac{T k_{B}^{2}}{\pi\hbar} \int_{0}^{\infty}\emph{d}\textrm{x}{\cal T}(\textrm{x}) \frac{\textrm{x}^{2}}{\sinh^{2}(\textrm{x})} \Delta T.
\label{conductance}
\end{equation}
where $T = (T_1+T_2)/2$ is the average temperature and $K$ is the thermal conductance.

Furthermore, if the two impedances 1 and 2 are two identical resistances, so that $R_1=R_2=R$, the photon transmission coefficient ${\cal T} =1$ and the thermal conductance $K$ is given by its quantized value, $K=K_Q= k_{B}^{2}T\pi/6 \hbar$. For mismatched resistances $R_1 \ne R_2$, the photon transmission coefficient ${\cal T}$ is smaller than 1 and the thermal conductance $K$ will be less than $K_Q$.

\subsection{Heat exchange with reactive coupling}

Let us now consider the more general case of circuit A with a non-zero reactive coupling impedance, following the same approach. Since no noise is generated in a purely reactive element, the spectral noise current through it is simply given by
\begin{equation}
\Delta I_c=\Delta V_c/Z_c.
\end{equation}
Imposing the constraint on  $\Delta I_i$ we then find $\Delta I(\omega) = [Z_1(\omega) \delta I_1(\omega) + Z_2(\omega) \delta I_2(\omega)]/Z_t(\omega)$. Now $Z_t(\omega)$ includes the coupling impedance, $Z_t(\omega) = Z_1(\omega)+Z_2(\omega)+Z_c(\omega)$. The constraint on  $\Delta V_i$ leads to $\Delta V_1(\omega) + \Delta V_2(\omega) = - Z_c(\omega) \Delta I(\omega)$. Moreover, $Z_1(\omega) \Delta V_2(\omega) - Z_2(\omega) \Delta V_1(\omega) = Z_1 (\omega)Z_2(\omega)[\delta I_1(\omega) -\delta I_2(\omega)]$. This system of equations can be solved for $\Delta V_1$ and $\Delta V_2$. The purely reactive impedance will neither emit nor absorb any power, we therefore obtain
\begin{equation}
P_1(t) = \langle \Delta I_{1} \Delta V_{1}\rangle = - \langle \Delta I_{2} \Delta V_{2}\rangle = -P_2(t).
\end{equation}
A straightforward calculation then yields Eqs.~(\ref{centrrescoef}) and (\ref{centrressimpl}), with the appropriate re-definition of the total series impedance $Z_t=Z_1 + Z_2 + Z_c$. This result generalizes the one presented in Ref.~\onlinecite{Nature-Meschke}.

\subsection{Heat exchange with mutual inductive coupling}

In this subsection, we consider configuration B, see Fig.~\ref{Circuits}. The mutual inductance $M$ relates the current fluctuations $\Delta I_c$ in the central coupling circuit to the voltage fluctuations in the two outer circuits, $\Delta V_{1,2}(\omega) = i\omega M \Delta I_c(\omega)$, hence we have $\Delta I_i(\omega) = \delta I_i(\omega) + i\omega M \Delta I_c(\omega)/Z_i(\omega)$. On the other hand, the current fluctuations $\Delta I_1$ and $\Delta I_2$ in the outer circuits are related to the fluctuation $\Delta I_c$ in the central circuit, according to the relation $\Delta I_c(\omega) =
i\omega M[\Delta I_1(\omega) + \Delta I_2(\omega)]/Z_c(\omega)$, where $Z_c(\omega)$ is the total series impedance of the elements in the central circuit, which we assume to be entirely reactive. Combining these observations, we conclude that
\begin{equation}
\Delta I_c(\omega) = \frac{i\omega M[\delta I_1(\omega) + \delta
I_2(\omega)]}{Z_c(\omega) + \omega^2 M^2 [1/Z_1(\omega_1) + 1/Z(\omega_2)]}.
\end{equation}
We substitute this result into the expressions for $\Delta I_{i}(\omega)$ and  $\Delta V_{i}(\omega')$, multiply them and obtain the power absorbed by the impedance $Z_i$ upon the appropriate Fourier transformation according to Eq.~(\ref{Pi}). Using the fact that $Z_c(\omega)$ is purely imaginary, we again find Eqs.~(\ref{centrrescoef}) and (\ref{centrressimpl}) to hold, but with the total impedance given by:
\begin{equation}
Z_t(\omega) = Z_1(\omega)+Z_2(\omega) + Z_c(\omega) \frac{Z_1(\omega) Z_2(\omega)}{\omega^2 M^2}.
\end{equation}
This is in agreement with the result obtained in Ref.~\onlinecite{PRL-Jauho}. However, we wish to note it is obtained here from quite simple circuit considerations, without the need of a Green function formalism.

\subsection{Heat exchange through a transmission line}

Let us finally study case C, see Fig.~\ref{Circuits}. A transmission line is represented by a series of N cells, each composed by an inductance L/N and a capacitance C/N. Considering again a photon wavelength larger than the size of a typical circuit element, we apply Kirchhoff's law locally and find a recurrence equation between the potential $V_n$ at element $n$ and the ones for the neighboring cells $V_{n-1}$ and $V_{n+1}$:
\begin{equation}
V_{n-1}+V_{n+1}+ \left( \omega^{2}\frac{LC}{N^{2}}-2 \right) V_{n}=0.
\end{equation}
Writing the local voltage as a plane wave $V_{n} \propto e^{i(kn-\omega t)}$ we obtain the dispersion relation
\begin{equation}
\cos k=1-\frac{\omega^{2}LC}{2N^{2}},
\end{equation}
describing propagating waves with a wave-vector dependent group velocity, thus taking into account retardation effects in the line.
Writing the expression of the current at both extremities of the transmission line we find
\begin{eqnarray}
\Delta I_{1}= -\frac{\Delta V_1}{Z_A}+\frac{\Delta V_2}{Z_B}, \\
\Delta I_{2}= -\frac{\Delta V_2}{Z_A}+\frac{\Delta V_1}{Z_B},
\end{eqnarray}
with the characteristic line impedances $Z_A$ and $Z_B$ defined as
\begin{eqnarray}
Z_{A}&=&\frac{i\omega L}{N(1-\cos k+\sin k \cot kN)},\\
Z_{B}&=&-i\omega \frac{L}{N} \frac{\sin kN}{\sin k}.
\end{eqnarray}
Identifying the above expressions with Eq.~(\ref{defin}) enables us to write the voltages $V_{1}$ et $V_{2}$ at the resistors 1,2 as a function of the intrinsic current fluctuations $\delta I_{1}$ et $\delta I_{2}$:
\begin{eqnarray}
\Delta V_{2}=\frac{Z_{B}^{2}\tilde{Z}_{2}}{\tilde{Z}_{1}\tilde{Z}_{2}-Z_{B}^{2}} \delta I_{2} + \frac{Z_{B}\tilde{Z}_{1}\tilde{Z}_{2}}{\tilde{Z}_{1}\tilde{Z}_{2}-Z_{B}^{2}} \delta I_{1},\\
\Delta V_{1}=\frac{Z_{B}^{2}\tilde{Z}_{1}}{\tilde{Z}_{1}\tilde{Z}_{2}-Z_{B}^{2}}\delta I_{1}+\frac{Z_{B}\tilde{Z}_{1}\tilde{Z}_{2}}{\tilde{Z}_{1}\tilde{Z}_{2}-Z_{B}^{2}}\delta I_{2},
\end{eqnarray}
where we have defined
\begin{equation}
\frac{1}{\tilde{Z}_{1,2}}=\frac{1}{Z_{1,2}}+\frac{1}{Z_{A}}.
\end{equation}
Using Eq.~(\ref{Pi}) while writing the quantities $\Delta V_{i} (\omega)$ and $\Delta I_{i}(\omega)$ in terms of the current fluctuations $\delta I_{i}$   gives the expression of the power absorbed by the impedance $Z_i$. We again obtain Eqs.~(\ref{centrrescoef}) and (\ref{centrressimpl}) to hold, with the total impedance given by:
\begin{eqnarray}
\nonumber Z_t(\omega)= \textit{i}\omega \frac{L}{N} \frac{\sin kN}{\sin k}+\frac{\cos k(N-1/2)}{\cos k/2}\left[Z_{1}+Z_{2}\right]\\
+\frac{Z_1 Z_2 N}{\textit{i} \omega L}\left[\left(\frac{\cos k(N-1/2)}{\cos k/2}\right)^{2}-1\right]\frac{\sin k}{\sin kN}.
\end{eqnarray}

\section{Results}

\subsection{Direct inductive coupling}

To be specific, let us first consider the case of circuit A with an inductive coupling, $Z_c = i \omega L$, coupling two identical resistors, $Z_1 = Z_2 = R$, kept at a small temperature difference $\Delta T$. Eq. (\ref{conductance}) then provides the expression for the thermal conductance $K_L$ upon substitution of
\begin{equation}
{\cal T}_L(\textrm{x}) = \frac{\alpha_L^{2}}{\alpha_L^{2}+\textrm{x}^{2}},
\label{filter2trans}
\end{equation}
which is the relevant photon transmission coefficient as a function of the dimensionless frequency $\textrm{x} = \hbar \omega/2 k_B T$. The parameter
\begin{equation}
\alpha_L=\frac{\hbar R}{L k_{B} T}
\end{equation}
defines the crossover between the low-frequency regime $\textrm{x} \ll \alpha_L$ where the inductance is transparent and the corresponding photon transmission coefficient ${\cal T}_L(\textrm{x})$ is close to 1 and the high-frequency regime $\textrm{x} \gg \alpha_L$ where the inductance becomes opaque and ${\cal T}_L(\textrm{x}) \sim \alpha_L^{2}/\textrm{x}^2 \ll 1$. The result (\ref{filter2trans}) for the photon transmission coefficient can thus be understood as originating from a low-pass LR filter composed of the resistance $R$ and the coupling inductance $L$ that filters the photonic thermal spectrum. The cut-off frequency is given by $\alpha_L$ in units of the thermal frequency $2k_BT/h$, at frequencies beyond the cut-off frequency the transmission decays as $1/\textrm{x}^2$.

\begin{figure}[t]
\centering
\includegraphics [width=8.5 cm]{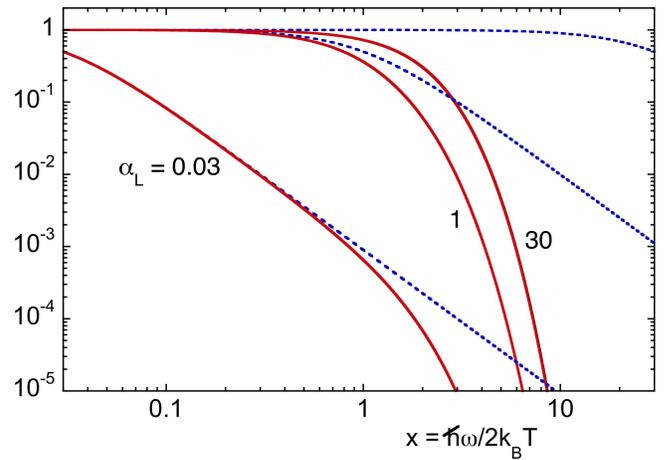}
\caption{(Color online) Case of an inductive coupling. Spectrum of the thermal noise power density $x^2/\sinh^2 x$ (black dots), the photon transmission coefficient ${\cal T}_L(x) = \alpha_L^2/(\alpha_L^2 + x^2)$ (dotted blue line) and of the product of the two, namely the photonic heat (thin red full line) as a function of the frequency, for values of the parameter $\alpha_L$ = 0.03, 1 and 30 from top to bottom. The frequency is plotted in units of the thermal frequency $2k_BT/\hbar$. We consider the case of perfect resistance matching $R_1 = R_2 = R$.}
\label{PlotsInductance}
\end{figure}

We have calculated the spectral density of the photonic heat transferred from one resistor to the other for several values of the parameter $\alpha_L$. Fig. \ref{PlotsInductance} displays the spectrum of the thermal noise current $\textrm{x}^{2}/\sinh^{2}(\textrm{x})$, the photon transmission coefficient ${\cal T}_L(\textrm{x})$ and of the product of the two. The integral of the latter quantity gives the power transmitted by photons $P_{1,2}$. In the limit $\alpha_L \gg 1$ of a small inductance, i.e. a negligible coupling impedance, the photon transmission coefficient equals to unity over the whole thermal spectrum. The integral is then equal to $\pi^2/6$ and one again recovers the quantum of conductance, $K_L = K_Q$. In the opposite limit $\alpha_L \ll 1$ of a large inductance, the photon transmission coefficient decays when the frequency is increased. The photonic signal is then strongly suppressed: $K_L \ll K_Q$.

\subsection{Direct capacitive coupling}

Let us now consider the case of circuit A with a capacitive coupling element $Z_c=1/i \omega C$. We assume again that the two impedances 1 and 2 are pure resistors with a small temperature difference. The thermal conductance $K_C$ is given by Eq. (\ref{conductance}) taking into account the photon transmission coefficient:
\begin{equation}
{\cal T}_C(\textrm{x}) = \frac{\textrm{x}^{2}}{\textrm{x}^{2}+\alpha_C^{2}}.
\label{filtertrans}
\end{equation}
The cross-over frequency is now determined by the parameter
\begin{equation}
\alpha_C=\frac{\hbar}{4 R C k_{B} T};
\end{equation}
it separates a low-frequency regime $\textrm{x} \ll \alpha_C$ where the capacitor is opaque and ${\cal T}_C(\textrm{x}) \sim \textrm{x}^2/\alpha_C^{2} \ll 1$ from a high-frequency regime $\textrm{x} \gg \alpha_C$ where the capacitor is transparent and ${\cal T}_C(\textrm{x}) \sim 1$.

\begin{figure}[t]
\centering
\includegraphics [width=8.5 cm]{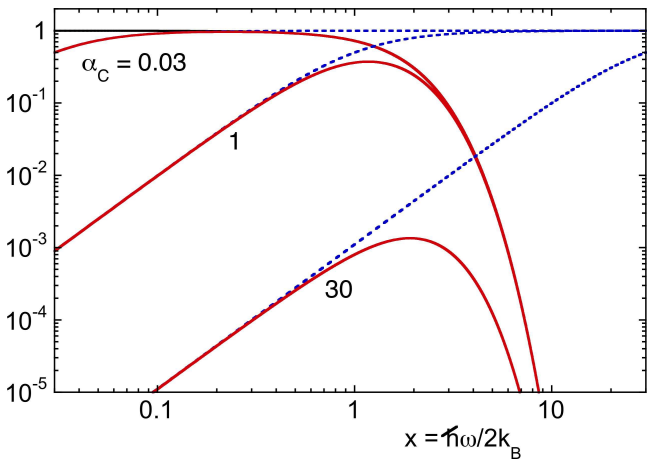}
\caption{(Color online) Case of a capacitive coupling. Spectrum of the thermal noise power density $x^2/\sinh^2 x$ (black dots), the photon transmission coefficient ${\cal T_C}(x) = x^2/(x^2+\alpha_C^2)$ (dotted blue line) and of photonic heat (thin red full line) as a function of the frequency, for values of the parameter $\alpha_C$ = 0.03, 1 and 30 from top to bottom. The frequency is plotted in units of the thermal frequency $2k_BT/\hbar$. We consider the case of perfect resistance matching $R_1 = R_2 =R$.}
\label{PlotsCapa}
\end{figure}

Fig.~\ref{PlotsCapa} displays information similar to that of Fig.~\ref{PlotsInductance} but for the case of a capacitance coupling the two resistors, yielding the photon transmission coefficient ${\cal T}_C$, Eq.~(\ref{filtertrans}). The limit $\alpha_C \ll 1$ means that the capacitance is large, i.e. it has a negligible impedance over most of the thermal spectrum. The transparency ${\cal T}_C$ is then equal to unity and one recovers $K_C = K_Q$. In the limit $\alpha_C \gg 1$, the photonic signal is strongly suppressed by the RC filter composed of the series capacitance and the receiver resistance,
leading to $K_C \ll K_Q$.

\subsection{Mutual coupling to an LC-resonator}

We now turn to circuit B, for the case where the central coupling circuit is an LC-resonator, as indicated in Fig.~\ref{Circuits}. This means that $Z_c(\omega) = i\omega L + 1/i\omega C$, the resonant frequency is given by $\omega_0 = \sqrt{1/LC}$. Assuming again $R_1=R_2=R$ and a small temperature difference $\Delta T$, the thermal conductance $K_M$ is given by Eq. \ref{conductance} with the photon transmission coefficient:
\begin{equation}
{\cal T}_M(\textrm{x}) = \frac{1}{1 + \alpha_M^2 (\textrm{x}/\gamma - \gamma/\textrm{x})^2 \textrm{x}^{-4}}.
\label{filterMtrans}
\end{equation}
Here
\begin{equation}
\alpha_M=\frac{R}{2} \sqrt{\frac{C}{L}} \left(\frac{L}{M}\right)^2 \gamma^2
\end{equation}
and
\begin{equation}
\gamma = \hbar \omega_0/2 k_B T.
\end{equation}
The photon transmission coefficient is characterized by a resonance at $\omega_0$, the width of which is governed by the parameter $\alpha_M$: the larger $\alpha_M$, the narrower the resonance. For small values of $\alpha_M$, the transmission coefficient is close to 1 over the thermal spectrum frequency range and we find $K_M \alt K_Q$.

\begin{figure}[t]
\centering
\includegraphics [width=8.5 cm]{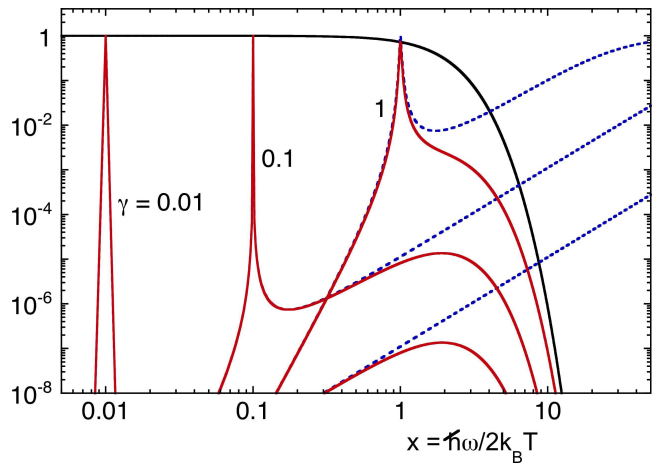}
\caption{(Color online) Case of a mutual coupling via an LC-resonator circuit. Spectrum of the thermal noise power density $x^2/\sinh^2 x$ (black dots), the spectral transmission factor ${\cal T}_M(x)$ (dotted blue line) and of the photonic heat (thin red full line) as a function of the frequency, for values of the parameter $\gamma$ = 0.01, 0.1 and 1. The frequency is plotted in units of the thermal frequency $2k_BT/\hbar$. We consider the case of perfect resistance matching $R_1 = R_2$ and a parameter $\alpha_M = 30$.}
\label{PlotsRes}
\end{figure}

\subsection{Coupling through a transmission line}

We finally deal with circuit C, with two pure resistors $R_1$ and $R_2$ separated by a transmission line that behaves like a low pass filter with a cut-off frequency $\omega_{c}=2/\sqrt{L C}$. In the limit of very low frequency $\omega \ll \omega_{c}$, the total impedance ${Z}_{t}$ is
\begin{eqnarray}
\nonumber \emph{Z}_{t}(\textrm{x})=(R_{1}+R_{2})\cos(N\alpha_{TL}\textrm{x})+ \\
\textit{i}\sin(N\alpha_{TL}\textrm{x})\sqrt{\frac{L}{C}}\left[1+R_{1}R_{2}\frac{C}{L}\right],
\end{eqnarray}
where
\begin{equation}
\alpha_{TL}=4 k_B T/\hbar \omega_c.
\end{equation}
For frequencies below $\omega_{c}$, the impedance $Z_t$ features a series of resonances. At every resonance including the zero-frequency case, the transmission line is fully transparent and the impedance $Z_t$ is equal to $R_1+R_2$.

At frequencies well above the cut-off frequency $\omega \gg \omega_c$, we can make the approximation $\cos k \simeq -\omega^{2}LC/2N^{2}$. The wave-vector k is then complex. The total impedance is purely imaginary and diverging at high frequency as:
\begin{equation}
\emph{Z}_{t}=\textit{i}\alpha_{TL}\textrm{x} \sqrt{\frac{L}{C}} [-\textrm{x}^2 \alpha_{TL}^{2}]^{N-1}.
\end{equation}

\begin{figure}[t]
\centering
\includegraphics [width=8.5 cm]{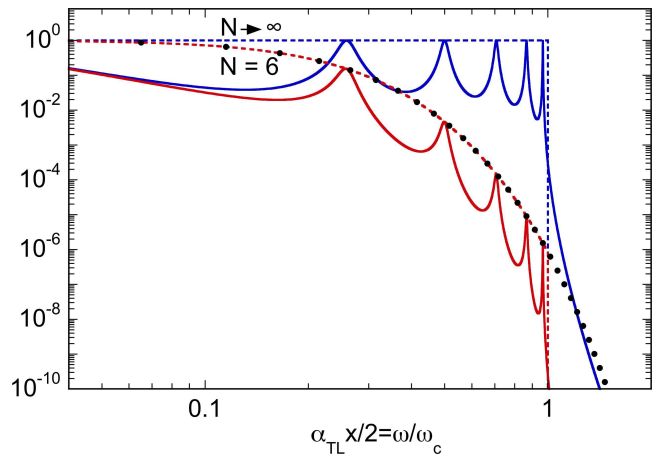}
\caption{(Color online) Case of a coupling through a transmission line. Spectrum of the thermal noise power density $x^2/\sinh^2 x$ (black dots), the photon transmission coefficient ${\cal T}_{TL}(x)$ (dotted and full blue lines) and of the photonic heat (dotted and full red lines) as a function of the frequency. The full lines stand for a number of cells $N =$ 6 and the dotted lines for $N$ infinite. The frequency is plotted in units of the cut-off frequency $\omega_c$. We consider a temperature T so that $\alpha_{TL} =$ 0.2 and the case of perfect resistance matching $R_1 = R_2$ with $R = 0.1 \sqrt{L/C}$.}
\label{PlotsLigne}
\end{figure}

From Eq. \ref{centrrescoef}, the transmission coefficient ${\cal T}_{TL}$ is related to the line impedance $Z_t$ as: ${\cal T}_{TL}(\omega) = 4 R_1 R_2/\mid Z_{t}(\omega)\mid^{2}$. Fig. \ref{PlotsLigne} shows the transmission coefficient for a number of cells $N =$ 6 and $N = \infty$, in the case $R_1=R_2 = 0.1 \sqrt{L/C}$, $\alpha_{TL} =$ 0.2. For a finite $N =$ 6, we observe a series of resonance peaks, featuring a maximum transmission equal to unity. The case $N = \infty$ corresponds to the continuum limit, where the discretization necessary for the calculation vanishes. In this case, the impedance $Z_t$ is constant and equal to $R_1+R_2$ over the full bandwidth $[0,\omega_c]$ and infinite above. The transmission is then unity within the bandwidth (if $R_1=R_2$) and zero above.

\subsection{Total photonic power with reactive coupling}

\begin{figure}
\centering
\includegraphics[width=8.5 cm]{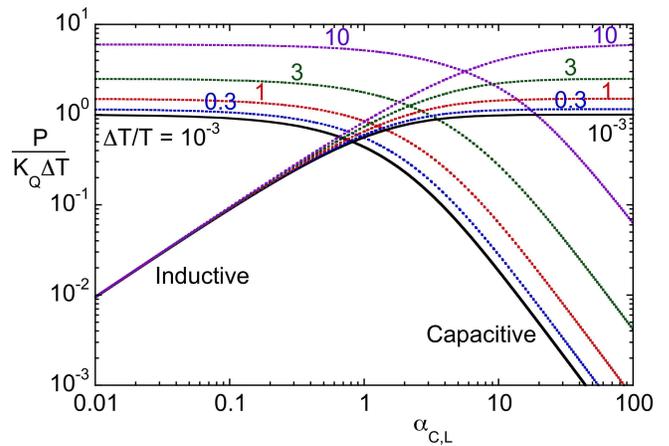}
\caption{Dependence of the photonic power through a capacitive or inductive coupling impedance on the parameters $\alpha_C$ or $\alpha_L$ for different values of the relative temperature difference $\Delta T/T$, in units of $K_Q \Delta T$, the maximum photonic power in the case of a linear response.}
\label{TotalPowerLC}
\end{figure}

Finally, we have calculated the total photonic power, integrated over the full frequency range, as a function of the parameter $\alpha_L$ or $\alpha_C$  in the respective cases of an inductive or capacitive coupling. We compare both linear and nonlinear response, changing the values of the relative temperature difference $\Delta T/T$, see Fig.~\ref{TotalPowerLC}. In the limit of a small temperature difference and with an inductive coupling, the photonic power decays as $\alpha_L$ when $\alpha_L$ is small. For a capacitive or mutual coupling, the total power is maximal for small $\alpha_{C,M}$; it decays as  $1/\alpha_{C,M}^2$ when $\alpha_{C,M}$ is large.

A cross-over between the linear regime $P \propto \Delta T$ and the non-linear regime occurs at $\Delta T/T \approx 1$. When the temperature difference is large, the photonic thermal conductance is larger than the quanta $K_Q$, because of the broader frequency range of the emitted photons. Only in the case of a significantly inductive coupling $\alpha_L > 1$, which cuts the high frequencies induced by the higher source temperature, does the thermal conductance not depend on the temperature difference.

In the cases of a coupling through a resonator or a transmission line, the behavior of the total photonic power as a function of the relevant parameter $\alpha_M$ or $\alpha_{TL}$ is similar to the capacitive case.

\section{Conclusion}

In summary, we have introduced a simple circuit approach of the photonic heat transport that can be applied to a variety of experimentally-relevant situations. We would like to stress that this approach operates within a very simple formalism, thus providing an intuitive understanding of photonic thermal conduction channel. This approach enabled us to investigate the phonic heat transfer through a reactive transmission line.

To conclude, let us discuss the possibility to practically use the photonic channel discussed above to transmit power between two electronic circuits that are galvanically isolated. This can be achieved by coupling two resistors through two capacitances. From the above discussion, the maximum efficiency can be attained when the source and receiver impedance are well matched and with a parameter $\alpha_C$ below about 0.1. With impedances of 50 $\Omega$, this corresponds to a capacitance value larger than 0.1 pF at a temperature of 4 K or 10 pF at 40 mK, which is significantly higher that the 1-10 fF capacitance of typical submicron scale junctions. In Ref. \onlinecite{PRL-Timofeev-1}, the estimated 10 fF ground capacitance gives a $\alpha_C$ parameter value of about 38 at a temperature of 100 mK. From Fig. \ref{TotalPowerLC}, one extracts that this results in a heat transfer rate divided by about 1000 compared to the full value, which is consistent with experimental results.\cite{PRL-Timofeev-1} In a real thermal circuit, large series capacitances can easily be integrated using available microfabrication technologies and ensure a fully-efficient photonic channel thermal coupling, while maintaining galvanic insulation. The related thermal conductance is about 1 pW/K at 1 K and scales linearly with the temperature. This approach is thus compatible with the cooling of small objects well decoupled from the thermal bath, like for instance membranes .

As discussed above, photonic heat can also be transmitted through a transmission line, provided that its cut-off frequency is over the thermal spectrum bandwidth. Let is consider for instance a superconducting transmission line made of two strips of width 1 $\mu$m, thickness 50 nm, and separated by 2 $\mu$m. Its kinetic inductance is about 10 pH/mm and its capacitance to the ground about 0.2 fF/mm. The related cut-off frequency is then about 1 K for a length of 10 mm.

Finally, it is interesting to discuss the present results in connection with the "brownian refrigeration" of a cold normal metal in contact with a superconductor via a capacitively shunted tunnel barrier, subjected to the thermal noise generated in a hot resistor~\cite{PRL-Brownian}. For typical system parameters, an optimum exists where a hot resistor of resistance $R = 10 R_K$ with $R_K = h/e^2$ gives rise to heat extraction currents of about $10^{-3} \Delta^2 /e^2 R_T$, where $\Delta$ is the superconducting gap and $R_T$ is the tunnel barrier normal-state resistance. However, at the same time, the photon heat channel discussed in the present work will be active, inducing a heat current from the hot resistor {\em towards} the cold normal metal. This was not included in Ref.~\onlinecite{PRL-Brownian}, where the resistance $R_N$ of the normal metal was set to zero. We can estimate the effect of the photon heat current by modeling the set-up as in A-type circuit: a series combination of the hot and cold resistors of resistance $R$ and $R_N$, respectively, coupled by a capacitance $C$. Using typical parameters of Ref.~\onlinecite{PRL-Brownian} and assuming a resistance mismatch $R_N/R =$ 0.01, the direct photon heat current can be estimated to be $10^{-3} \Delta^2 /e^2 R_K$, which is comparable to or larger than the heat extraction current if $R_T \agt R_K$. In order to recover a net cooling of the cold resistor, it is not possible to reduce $R_T$ below $R_K$, as spurious higher order tunneling processes in the junction would reduce the heat extraction current. One can however increase the resistance mismatch: the photon heat current is proportional to $R_N/R$. We conclude that the heat extraction mechanism of Ref.~\onlinecite{PRL-Brownian} outweighs possible direct photon heating only for metals with resistances $R_N$ of a few Ohms or less.

This work is supported by the R\'egion Rh\^one-Alpes, the ANR project 'Elec-EPR' and the NanoSciERA project 'Nanofridge'. We thank C. Urbina for raising the question that motivated this work, O.-P. Saira and J. P. Pekola for discussion.

\end{document}